\documentclass[12pt]{article}

\usepackage{latexsym}
\usepackage{amsmath}
\usepackage{color}
\usepackage{adjustbox}
\usepackage{natbib}
\setcitestyle{citesep={;}, aysep={,}}

\usepackage{tikz}
\usepackage{float}
\usepackage{caption} 
\def\checkmark{\tikz\fill[scale=0.4](0,.35) -- (.25,0) -- (1,.7) -- (.25,.15) -- cycle;}

\numberwithin{equation}{section}

\newtheorem{theorem}{Theorem}

\newtheorem{example}{Example}[section]

\newenvironment{proof}[1][Proof]{\textbf{#1.} }{\
\qquad\qquad\rule{0.5em}{0.5em }}

\newcommand{\blind}{1}

\begin{document}

\def\spacingset#1{\renewcommand{\baselinestretch}%
{#1}\small\normalsize} \spacingset{1}


\if1\blind
{
  \title{\bf A-Optimal Split Questionnaire Designs for Multivariate Continuous Variables}
  \author{Dae-Gyu Jang\thanks{
    The authors gratefully acknowledge \textit{USDA NRCS Cooperative Agreement, Grant/Award Number: 68-7482-17-009;}}\hspace{.2cm}\\
    Department of Statistics, Iowa State University\\
    Zhengyuan Zhu \\
    Department of Statistics, Iowa State University
    and \\
        Cindy Yu \\
    Department of Statistics, Iowa State University\\
    }
  \maketitle
} \fi

\if0\blind
{
  \bigskip
  \bigskip
  \bigskip
  \begin{center}
    {\LARGE\bf A-Optimal Split Questionnaire Designs for Multivariate Continuous Variables}
\end{center}
  \medskip
} \fi

\bigskip
\begin{abstract}
A split questionnaire design (SQD), an alternative to full questionnaires, can reduce the response burden and improve survey quality. One can design a split questionnaire to reduce the information loss from missing data induced by the split questionnaire.  This study develops a methodology for finding optimal SQD (OSQD) for multivariate continuous variables, applying a probabilistic design and optimality criterion approach. Our method employs previous survey data to compute the Fisher information matrix and A-optimality criterion to find OSQD for the current survey study. We derive theoretical findings on the relationship between the correlation structure and OSQD and the robustness of local OSQD. We conduct simulation studies to compare local and two global OSQDs; mini-max OSQD and Bayes OSQD) to baselines. We also apply our method to the 2016 Pet Demographic Survey (PDS) data. In both simulation studies and the real data application, local and global OSQDs outperform the baselines.
\end{abstract}

\noindent%
\vfill

\newpage
\spacingset{1.45} 
\section{Introduction}

Survey sampling is the main method for collecting data in many disciplines. 
Researchers often need to collect data on hundreds of variables in order to make inferences on many population parameters and study their relationships. The need for detailed information leads to  lengthy questionnaires, which increases the response burden and has the potential to damage the survey response quality. Studies have found that survey response rates are lower when the questionnaire's length is longer \citep{galesic2009effects, rolstad2011response}. In addition, excessive questionnaire length may result in survey fatigue and reduced accuracy in the responses \citep{galesic2009effects, gibson2019effects}. To address this concern, researchers sometimes use a split questionnaire design (SQD), a survey design that splits a lengthy questionnaire into subsets of questions and assigns a subset to a respondent. The use of SQD shortens the survey length, which reduces the response burden and can improve the response quality.

The use of split questionnaires only allows us to observe a subset of variables for each respondent, which creates missingness in the resulting survey data. It is called missing by design. In the analysis of incomplete data, the design of the missing pattern affects the properties of estimators calculated based on SQDs \citep{thomas1997generating, rhemtulla2016asymptotic}. Therefore how to build an SQD that renders the most efficient and accurate estimates of population parameters has been an interesting and important question to researchers.

In early studies, simple and heuristic SQD methods were used. \cite{shoemaker1973physiologic} used multiple matrix sampling, which involves the randomized selection of both respondents and question items. \cite{raghunathan1995split} used the partial correlation coefficients of variables to assign the variables with high partial correlation coefficients to different subsets. The above two methods are intuitively appealing and useful, but they did not provide the framework to maximize the amount of information from partially observed survey data since they do not employ an optimization procedure.

To address this issue, researchers have proposed methods to find an optimal SQD based on different objective functions. \cite{Thomas2006AnEO} proposed an automatic method for creating subsets of question items in a way that items included in a subset are predictive of excluded items. They developed an index of predictive value, which estimates the contribution of a subset of items to the estimation of the means of the excluded items. This method aims to find an optimal design that minimizes the variances of multiple imputation estimators. Another approach is to choose among several predetermined choices a design that minimizes the Kullback-Leibler (KL) divergence, which measures the amount of information loss between an SQD data and a full questionnaire data \citep{adiguzel2008split, stuart2019computationally}. The optimal design based on the KL divergence yields smaller variances of estimators than a heuristic design. \cite{gonzalez2008adaptive} proposed an adaptive matrix sampling method that uses the first interview information to assign probabilities of subsampling items for the second interview. They used optimality criteria in the evaluation and comparison of five predetermined allocation methods. They found that the allocation method that assigns the subsampling probabilities proportional to the absolute relative mean deviation is optimal among the five methods. However, they did not consider finding an optimal allocation method in a more general design space. \cite{chipperfield2009design, chipperfield2011efficiency} discussed choosing the optimal SQD that minimizes a variance function for a fixed cost or minimizes the survey cost for a fixed variance. Their method determines the optimal sample size allocation to each SQD pattern, where an SQD pattern refers to a way of assigning question items. \cite{zhang2019active} proposed the active question selection method that chooses questions sequentially by minimizing the Bayes A-optimality, the sum of posterior variances of latent variables called user factors.

 As seen above, different researchers have proposed different objective functions for finding the optimal SQD, but the use of optimality criteria has not been discussed extensively other than in \cite{gonzalez2008adaptive} and \cite{zhang2019active}. However, \cite{gonzalez2008adaptive} did not seek an optimal SQD in the set of all feasible designs and only compared several predetermined choices. The proposed method in \cite{zhang2019active} obtained a deterministic question order for all respondents provided with a Gaussian response model. All respondents are provided the same question items, which is unrealistic in surveys. Optimality criteria have not been commonly used in the search for an optimal SQD from a large set of designs partially due to computation complexity, even though they are popular in the area of experimental designs. The presence of missing values in SQD data makes it challenging to estimate the covariance matrix of estimators, which is needed as an argument for optimality criteria. Methods for estimating the covariance matrix of estimators in SQD have been presented in several recent papers, such as \citep{kim2012factoring, chipperfield2011efficiency, rhemtulla2016asymptotic}. Thanks to these studies, it has become plausible to implement optimality criteria in SQD.

In this article, we develop a rigorous mathematical foundation for general SQDs and define an optimal SQD (OSQD) based on optimal experimental design and probability sampling theories. We view the selection of a set of questions as a random sampling of question items from the complete questionnaire. Therefore, an SQD is a sampling design with a probability distribution on the space of all possible samples of questions. The majority of other methods only consider a small number of split questionnaire patterns and optimize the proportions of the split questionnaire patterns in the sample. We consider a much larger space of possible split questionnaire patterns by treating the selection of questions for each respondent as a separate sampling problem. We thus propose a method to choose the OSQD as the sampling design that optimizes an optimality criterion where inclusion probabilities for the SQD are arguments for the objective function. We compare our method in simulation studies and real data applications. Our OSQD leads to a more efficient estimator than other baseline methods.

The rest of this paper is organized as follows. Section 2 introduces the new framework for SQD. Section 3 describes the concept of A-optimal SQD (A-OSQD), statistical models used as the super-population model for the survey response, and presents some theoretical results. Section 4 conducts simulation studies to evaluate the finite sample performance of the A-OSQD. Section 5 presents an application of our method to the 2016 Pet Demographic Survey (PDS) conducted by the American Veterinary Medical Association (AVMA). Section 6 states conclusions and discussions.

\section{New Framework for SQD}
SQD is a form of matrix sampling since it samples elements of a response matrix. We first introduce the notation and setup for matrix sampling to provide the mathematical foundation for general SQDs. Assume that there are $ K $ question items in a full questionnaire and $ N $ units in a population. We have a $N \times K$ matrix of responses. A resulting combination of sampled units and selected questions from conducting an SQD corresponds to an outcome from matrix sampling in which each element of the matrix is sampled randomly. For $k=1, 2, \dots, K$, and $i=1, 2, \dots, N$, the subsampling indicators are defined as $\alpha_{ik}=1$, if the $i$th unit is sampled from the population and the $k$th question item is chosen to be administered by the $i$th unit, 0 otherwise. For $i=1, \dots, N$, we use $\boldsymbol{Y_i}$ to denote the $K$-variate response vector for $i$th sample unit. Let $p_{ik}$ be the probability that $\alpha_{ik}=1$ for unit $i$ and item $k$, and $\{\bar{Y}_k\}_{k=1}^K$ be the population means of $K$ questions which are the parameters of interest. Then, the H{\'a}jek (HK) estimator of the population mean is written as follows:\\
\begin{equation} \label{eq:1}
\hat{\bar{y_k}}=\left(\sum_{i=1}^N \alpha_{ik}/p_{ik} \right)^{-1}\left(\sum_{i=1}^N \alpha_{ik} y_{ik}/p_{ik} \right),
\end{equation}
where $y_{ik}$ is the response of the $i$th unit and the $k$th item.

In a survey, we often sample $n$ respondents first, and then choose a subset of questions to be administered. In such a situation, $p_{ik}=p^{*}_{i} p^{**}_{k|i}$, where $p^{*}_{i}$ is the first order inclusion probability that unit $i$ is sampled in the first phase, and $p^{**}_{k|i}$ is the probability that question item $k$
is administered by unit $i$, given that unit $i$ is sampled. $p^{**}_{k|i}$ is related to our research question, that is, how to obtain the optimal selecting probabilities for SQD patterns. To focus on this research question, we assume that the cost of administering each question is the same, and $ n $ units are sampled by simple random sampling without replacement (SRS), i.e., \ $p^{*}_{i}=n/N$. The selection of question items can be viewed as choosing one from a set of SQD patterns, where an SQD pattern refers to an assignment of items to a split questionnaire. The number of all possible SQD patterns is $2^K-1$, which can become extremely large even for a moderate $K$. 
We next introduce a few additional constraints to make this a more manageable problem.
We denote $J$ as the number of SQD patterns considered.  In our setup, we additionally assume that the number of chosen questions for each sample unit is fixed as $m(=\sum_{k=1}^K \alpha_{ik})$. It is a reasonable assumption that ensures that the response burden for each unit is balanced. Under this assumption $J=\binom{K}{m}$.  Let $A$ be a set of indexes for question items included in an SQD pattern. For example, $A=\{1, 3\}$ is an index set for an SQD pattern that includes item 1 and item 3. Define $\pi_{A|i}$ as the probability that the SQD pattern $A$ is selected for unit $i$. Then, $p^{**}_{k|i}=\sum_{\{A: k \in A\}} \pi_{A|i}$. When the distribution of $\boldsymbol{Y_i}$ is significantly different in demographic subgroups defined by variables such as age, gender, and income, we may consider using this information to determine $\pi_{A|i}$ for different units. In this paper for simplicity we assume that $\pi_{A|i}$ is constant, i.e.\, $\pi_{A|i}=\pi_A$ for all $i$. Thus, $p^{**}_{k|i}$ does not depend on $i$, i.e.\, $p^{**}_{k|i}=p^{**}_{k}$ for each $k$ and all $i$. We will consider the more general setup in follow-up work.

 We will give a toy example to illustrate the setup more clearly. Consider a survey that has four questionnaires, we want to select ten units, and each unit answers two of the four questions. In this case,  $K=4$, $n=10$, $m = 2$, and $J=\binom{4}{2} =6 $. Table \ref{table:1} provides detailed information about SQD patterns and their associated probabilities.

\begin{table}[ht]
\caption{SQD patterns and their associated probabilities}
\begin{center}
    \begin{tabular}{llllll}
\hline
pattern & Q1                        & Q2 & Q3 & Q4 & prob. \\ \hline
1       & \checkmark &  \checkmark  &    &    &   $\pi_{\{1, 2\}}$    \\
2       &                           &    &   \checkmark &  \checkmark  &  $\pi_{\{3, 4\}}$    \\
3       &               \checkmark            &    & \checkmark   &    &   $ \pi_{\{1, 3\}}$  \\
4       &                           &  \checkmark  &    & \checkmark   &    $\pi_{\{2, 4\}} $  \\ 
5      &               \checkmark            &    &    & \checkmark   &    $\pi_{\{1, 4\}}$  \\
6       &                           &  \checkmark  &  \checkmark    &  &    $\pi_{\{2, 3\}}$   \\ \hline
\end{tabular}
\\ $\sum_{1 \le k < k` \le 4} \pi_{\{k, k`\}}=1.$
\end{center}

\label{table:1}
\end{table}
 
Under the assumptions above, $p_{ik} \propto p^{**}_{k}$ for each $k$ and all $i$. Suppose $n=10$ sample units are drawn using SRS. Table \ref{table:2} gives an example of chosen questions resulting from this SQD design.

\begin{table}[ht]
\caption{An example of chosen questions.}
\begin{center}
    \begin{tabular}{lllll}
\hline
Respondent & Q1 & Q2 & Q3 & Q4 \\ \hline
1          &  \checkmark  &  \checkmark  &    &    \\ 
2          &    &  \checkmark  &    &  \checkmark  \\ 
3          &  \checkmark  &    &  \checkmark  &    \\ 
4          &  \checkmark  &  \checkmark  &    &    \\ 
5          &    &    &   \checkmark &  \checkmark  \\ 
6          &    &  \checkmark  &    & \checkmark   \\
7          &  \checkmark  &    &    &  \checkmark  \\ 
8          &    &  \checkmark  &    & \checkmark   \\
9          &  \checkmark  & \checkmark   &    &    \\ 
10          &    &    &  \checkmark  & \checkmark   \\ \hline
\end{tabular}
\end{center}

\label{table:2}
\end{table}

Then, we have the HK estimator of $\bar{Y_3}$:
\begin{equation} \label{eq:2}
    \hat{\bar{y_3}} =\left(\sum_{i=1}^n \alpha_{i3}/p^{**}_{3} \right)^{-1}\left(\sum_{i=1}^n \alpha_{i3} y_{i3}/p^{**}_{3}\right)
   =\frac{1}{n_3} \sum_{i=1}^{10} \alpha_{i3}y_{i3},
\end{equation} which is the sample mean of the observed responses for question item 3, where $n_3=\sum_{i=1}^{10} \alpha_{i3}$. It can be shown that the variance of the above HK estimator is approximately equal to $(\frac{N}{np^{**}_k}-1)\sum_{i=1}^N (y_{ik}-\bar{y}_k)^2$, which indicates that the higher $p^{**}_k$ is assigned, the smaller variance of the estimator."

\section{A-OSQD}
In statistical inference, we specify a statistical model and construct a legitimate estimator to estimate the model parameters of interest. For many cases, the covariance matrix of estimators is known or can be computed. In a design of an optimal experiment, an optimal design can be defined as a design that minimizes or maximizes an optimality criterion, a function of the covariance matrix of estimators. Therefore, to define an OSQD using the optimal design method, we need to state a statistical model, estimators of parameters, and an optimality criterion.  In this study, we consider two models: the multivariate normal (MVN) model and the zero-inflated multivariate log-normal model (ZMVLN), and use maximum likelihood estimators (MLE) to estimate population means. We also employ the A-optimality criterion to minimize the sum of variances of estimators. The following subsections will introduce notation and terminology in OSQD,  two statistical models, and the A-optimality criterion based on the models.
\subsection{Optimality Criteria}
 Let $P:A \rightarrow \pi_A$ be the probability of choosing a SQD pattern, $A$. Let $\mathcal{P}$ be the set of all possible $P$. Then, a design of split questionnaire can be seen as a specification of $P\in \mathcal{P}$. We assume that the observation vector, $\boldsymbol{Y}$  has a density function $f(\boldsymbol{y}; \boldsymbol{\theta})$, where $ \boldsymbol{\theta}$ is the parameter. Denote $f_A(\boldsymbol{y}_A; \boldsymbol{\theta})$ be the marginal density function of $\boldsymbol{y}_A$, where $\boldsymbol{y}_A$ is the observed part of $\boldsymbol{y}$ with the set of chosen questions in $A$.  Let $\{A_j\}_{j=1}^J$ be the set of all SQD patterns prescribed in a SQD $P$. Then, $I(\boldsymbol{\theta}; P)$, the fisher information matrix of $\boldsymbol{\theta}$ given a SQD $P$, is written as follows:\\
\begin{equation} \label{eq:3}
    I(\boldsymbol{\theta}; P)=-\sum_{j=1}^J P(A_j) E\left[\frac{\partial^2 f_{A_j}(\boldsymbol{y}_{A_j}; \boldsymbol{\theta})}{\partial \boldsymbol{\theta}\partial\boldsymbol{\theta}^T}\right].
\end{equation}
Note that the inverse of the Fisher information matrix concerns the asymptotic variance of the MLE under a specified density.
 In this framework, $P_{opt}$, an OSQD, is written as follows:\\
\begin{equation} \label{eq:4}
    P_{opt}=arg \min_{P \in \mathcal{P}} \Psi\left(I(\boldsymbol{\theta}; P)\right),
\end{equation}
where $\Psi$ is a function that satisfies the following properties.

\begin{enumerate}
    \item Monotonicity\\
    If $M_1$ and $M_2$ are two information matrices such that $M_2-M_1$ is non-negative definite, then $\Psi(M_1) \le \Psi(M_2)$.
    \item Homogeneity\\
    $\Psi(rM)=\gamma(r) \Psi(M)$ for non-increasing function $\gamma$, any information matrix $M$, and all $r>0$. 
    \item Convexity\\
    For any $\alpha \in [0,1]$, and any information matrices $M_1$ and $M_2$, $\Psi(\alpha M_1 + (1-\alpha) M_2) \le \alpha \Psi(M_1)+(1-\alpha) \Psi(M_2)$.
\end{enumerate}
$\Psi\left(I(\boldsymbol{\theta}; P)\right)$ is called an optimality criterion. Here, we introduce a few popular optimality criteria.
\begin{enumerate}
    \item A-optimality: $\Psi(M)=tr(M^{-1})$ for the information matrix $M$, i.e.\ equation (\ref{eq:4}) minimizes the trace of the covariance matrix of the MLE.
    \item D-optimality: $\Psi(M)=(det(M))^{-1}$ for the information matrix $M$, i.e.\  (\ref{eq:4}) minimizes the (log) determinant of the covariance matrix of the MLE.
    \item T-optimality: $\Psi(M)=-tr(M)$ for the information matrix $M$, i.e.\ equation (\ref{eq:4}) maximizes the trace of the information matrix.
    \item E-optimality: equation (\ref{eq:4}) maximizes the minimum eigenvalue of the information matrix.
\end{enumerate}
 Because the sum of variances is one of the popular measures, we focus on the unweighted A-optimality criterion in this study. However, it is straightforward to derive a weighted optimality criterion as follows. When $L\boldsymbol{\theta}$, linear combinations of the parameters are of interest, the inverse of Fisher information matrix can be replaced by $LM^{-1}L^T$. By putting $L=\text{diag}(w_1^{1/2}, \dots, w_q^{1/2})$, we can create weighted optimality criteria, where $w_i$ for $i=1, \dots , q$ are weights for parameters. Parameters of greater interest should be assigned heavier weights. If you take $w_i=1/E(Y_i)$, you minimize the sum of coefficients of variation instead of variances.\\
 In practice, the parameter $\boldsymbol{\theta}$ is unknown. In survey studies, one can use data from a previous survey or a pilot study to estimate $\boldmath{\theta}$ and find an optimal design. But the optimal design based on a value of $\boldmath{\theta}$ is locally optimal, so the bias in the value of $\boldmath{\theta}$ can damage the performance of an optimal design.  When the preliminary estimate of $\boldsymbol{\theta}$ is highly biased or unavailable, we can consider global optimal designs; Bayesian design, or mini-max design. Bayesian design, a design that minimizes Bayesian design criterion based on a prior density, $f_{prior}(\boldsymbol{\theta})$, is as follow:
 \begin{equation} \label{eq:3.3}
    P^B_{opt}=\arg \min_{P \in \mathcal{P}} \int_{\boldsymbol{\Theta}} \Psi\left(I(\boldsymbol{\theta}; P)\right) f_{prior}(\boldsymbol{\theta})d\boldsymbol{\theta}.
\end{equation}
Alternatively, mini-max design, a design that minimizes the maximum of an optimality criterion, is as follow:
\begin{equation} \label{eq:3.4}
    P^M_{opt}=\arg \min_{P \in \mathcal{P}}  \max_{\boldsymbol{\theta} \in \boldsymbol{\Theta}}\Psi\left(I(\boldsymbol{\theta}; P)\right).
\end{equation}

\subsection{Multivariate Normal Model: Fisher Information and A-optimality Criterion}

For multivariate continuous variables, the MVN model is one of the most popular models. For the MVN model applied to SQD, parameters and the fisher information matrix are defined as following. Suppose $\boldsymbol{Y}\sim MVN_K(\boldsymbol{\mu}, \boldsymbol{\Sigma})$, where
	$\boldsymbol{\theta} = (\boldsymbol{\mu}, vech(\boldsymbol{\Sigma}))$, and $vech(\boldsymbol{\Sigma})$ is the vector of all upper-triangular elements in $\boldsymbol{\Sigma}$.
$D_K$ is a $K^2 \times(K(K+1)/2)$ duplication matrix, such that $D_K vech(\boldsymbol{\Sigma})=vec(\boldsymbol{\Sigma})$, where $vec(\boldsymbol{\Sigma})$ is a vector of all elements in $\boldsymbol{\Sigma}$.
Using (\ref{eq:2}) and the property of MVN distribution, the fisher information matrix of $\boldsymbol{\theta}$ is written as follows:
\begin{equation} \label{eq:5}
    	I(\boldsymbol{\theta};P)=\sum_{j=1}^J P(A_j)\begin{bmatrix}
    \boldsymbol{\tau}_j^T\boldsymbol{\Sigma}_j^{-1}\boldsymbol{\tau}_j  & \boldsymbol{0} \\
    \boldsymbol{0} & 0.5D_K^T(\boldsymbol{\tau}_j^T\boldsymbol{\Sigma}_j^{-1}\boldsymbol{\tau}_j\otimes \boldsymbol{\tau}_j^T\boldsymbol{\Sigma}_j^{-1}\boldsymbol{\tau}_j)D_K
    \end{bmatrix},
\end{equation}
	where $\otimes$ denotes the Kronecker's product, $\boldsymbol{\tau}_j$ is a $m \times K$ matrix obtained by removing, from the $K \times K$ identity matrix, those rows corresponding to missing variables in pattern $A_j$, $\boldsymbol{\Sigma}_j$ is a $m \times m$ sub-matrix of $\boldsymbol{\Sigma}$ for observed variables for $A_j$ \citep{yuan20005, savalei2010expected}.\\
	We suppose $\boldsymbol{\mu}$ is the only parameter of interest, whereas $\boldsymbol{\Sigma}$ is a nuisance parameter. Then, $A(\boldsymbol{\theta}, P)$, the A-optimality criterion, which is the trace of the asymptotic covariance matrix of the MLE of $\boldsymbol{\mu}$ is as follows:
	\begin{equation}  \label{eq:6}
	   A(\boldsymbol{\theta}, P)= tr\left\{\left[\sum_{j=1}^J P(A_j) \boldsymbol{\tau}_j^T\boldsymbol{\Sigma}_j^{-1}\boldsymbol{\tau}_j \right]^{-1} \right\}.
	\end{equation}

    $P_{A-opt}$, the A-OSQD is the minimizer of $A(\boldsymbol{\theta}, P)$. Since $A(\boldsymbol{\theta}, P)$ is a function of an unknown parameter $\boldsymbol{\Sigma}$, the searching for OSQD requires an estimate of $\boldsymbol{\Sigma}$. To address this concern, We additionally assume that we can use data from the pilot study to obtain a preliminary estimator of $\boldsymbol{\Sigma}$ and minimize the A-optimality criterion that uses the plugged-in estimator of $\boldsymbol{\Sigma}$. $P^B_{A-opt}$, the Bayesian A-OSQD is the minimizer of  $\int_{\boldsymbol{\Theta}} A(\boldsymbol{\theta}, P) f_{prior}(\boldsymbol{\theta}) d\boldsymbol{\theta}$, where $f_{prior}(\boldsymbol{\theta})$ is the prior density. $P^M_{A-opt}$, the mini-max A-OSQD is the minimizer of $\max_{\boldsymbol{\theta} \in \boldsymbol{\Theta}} A(\boldsymbol{\theta}, P)$.

\subsection{Zero-inflated Multivariate Log-normal Model}
2016 PDS data has several traits to make it questioning to employ the MVN model. First, the variables are non-negative and non-symmetric. Second, the variables have many zeros. To deal with these issues, we specified a zero-inflated log-normal model. Since our survey data has multiple variables, the multivariate version of the zero-inflated log-normal model is required. The univariate version of the model has been discussed in literature such as \cite{chen2010confidence}, but the multivariate version has been barely discussed yet. Thus, in this study, we define a multivariate version of the zero-inflated log-normal model. The key idea is to generate a random vector from multivariate log-normal (MVLN) distribution with mean $\boldsymbol{\mu}$ and variance $\boldsymbol{\Sigma}$, and multiply independent Bernoulli random variables to each element.
 Let $\boldsymbol{Y^*}=(Y^*_1, \dots, Y^*_K)^T\sim MVLN_K(\boldsymbol{\mu}, \boldsymbol{\Sigma})$ be a random vector from a K-variate log-normal distribution. For $k=1, \dots, K$, $Z_k$'s are independent random variables from Bernoulli($\lambda_k$). Let $\boldsymbol{Y}=(Y_1, \dots, Y_K)^T$ has a $K$-variate ZMVLN distribution, where $Y_k=Y^*_k Z_k$  for all $k$. Let $\boldsymbol{\lambda}=(\lambda_{1}, \dots, \lambda_{K})^T$. Then, the unknown parameter is $\boldsymbol{\theta} = (\boldsymbol{\lambda}, \boldsymbol{\mu}, vech(\boldsymbol{\Sigma}))$. $\boldsymbol{Y}$ has a density
    \begin{equation} \label{eq:3.3.1}
    f(\boldsymbol{Y}; \boldsymbol{\theta})=f_{nz}(\boldsymbol{Y_{nz}}; \boldsymbol{\mu}, \boldsymbol{\Sigma})\prod_{k=1}^K \lambda_{k}^{I(Y_k \ne 0)} (1-\lambda_{k})^{I(Y_k=0)}, 
    \end{equation}
    where $\boldsymbol{Y_{nz}}$ is a vector of all non-zero elements in $\boldsymbol{Y}$, and $f_{nz}$ is the marginal MVLN density of $\boldsymbol{Y_{nz}}$. In the presence of missing values due to SQD, let $\boldsymbol{Y}_{obs}$ be the observed part of $\boldsymbol{Y}$. Then $f_{obs}$, the marginal density of $\boldsymbol{Y}_{obs}$, is written as follows:
    \begin{equation} \label{eq:3.3.2}
    	f_{obs}(\boldsymbol{Y}_{obs};\boldsymbol{\theta})=f_{obs,nz}(\boldsymbol{Y}_{obs,nz}; \boldsymbol{\mu}, \boldsymbol{\Sigma})\prod_{k \in A} \lambda_{k}^{I(Y_k \ne 0)} (1-\lambda_{k})^{I(Y_k=0)}, 
    \end{equation}
    where $\boldsymbol{Y}_{obs, nz}$ is the non-zero observed elements in $\boldsymbol{Y}$,  $f_{obs,nz}$ is the marginal density of $\boldsymbol{Y}_{obs,nz}$, and $A$ is an index set of questions items for a given SQD pattern. Since the missing mechanism is missing completely at random, $f_{obs,nz}$ is a marginal MVLN density. The following example helps you understand the notations.
    
\begin{example}
    Let $\boldsymbol{Y}=(Y_1, Y_2)^T$ has ZMVLN distribution with parameter $(\boldsymbol{\lambda}, \boldsymbol{\mu}, \boldsymbol{\Sigma})$. Let $A_{obs, nz}=\{t \in \{1,2\}: Y_t \text{ is observed and non-zero.}\}$. Then, there are 4 cases as following:
\begin{equation} \label{eq:3.3.3}
f_{obs,nz}(\boldsymbol{Y}_{obs,nz}; \boldsymbol{\mu}, \boldsymbol{\Sigma}) = \left\{
  \begin{array}{ll}
    1, & \text{if } A_{obs,nz}=\emptyset, \\
    y_1^{-1} \phi_1((log(y_1)-\mu_1)/\Sigma_{11}), & \text{if } A_{obs,nz}=\{1\}, \\
    y_2^{-1} \phi_1((log(y_2)-\mu_2)/\Sigma_{22}), & \text{if } A_{obs,nz}=\{2\}, \\
    y_1^{-1} y_2^{-1} \phi_2(log(y_1), log(y_2); \boldsymbol{\mu}, \boldsymbol{\Sigma}), & \text{if } A_{obs,nz}=\{1, 2\},
  \end{array}
\right.
\end{equation}
where $\phi_1$ is the density function of $N(0, 1)$, and $\phi_2(\cdot, \cdot ;\boldsymbol{\mu}, \boldsymbol{\Sigma})$ is the density function of the bi-variate normal distribution with mean $\boldsymbol{\mu}$ and covariance matrix $\boldsymbol{\Sigma}$.
    
\end{example}

From the density function, the fisher information matrix can be derived as follows:
    \begin{equation} \label{eq:3.3.4}
    I(\boldsymbol{\theta}; P)=\begin{bmatrix}
\boldsymbol{B_1} & \boldsymbol{0} \\
\boldsymbol{0} & \boldsymbol{B_2}
\end{bmatrix},
    \end{equation}
    where $B_1=\textbf{diag}\left(\frac{\xi_1}{\lambda_1 (1-\lambda_1)}, \dots, \frac{\xi_K}{\lambda_K (1-\lambda_K)}\right)$, 
	\[
	B_2=\sum_{j=1}^J \pi_{A_j} \sum_{k=1}^{2^{|A_j|}} E_{j_k} \begin{bmatrix}
    \boldsymbol{\tau}_k^T\boldsymbol{\Sigma}_k^{-1}\boldsymbol{\tau}_k  & \boldsymbol{0} \\
    \boldsymbol{0} & 0.5D_K^T(\boldsymbol{\tau}_k^T\boldsymbol{\Sigma}_k^{-1}\boldsymbol{\tau}_k\otimes \boldsymbol{\tau}_k^T\boldsymbol{\Sigma}_k^{-1}\boldsymbol{\tau}_k)D_K
    \end{bmatrix},
	\]
	where $\xi_i=\sum_{\{\text{all} ~A~ \text{containing} ~i\}} \pi_A$, $\{E_{j_k} : k=1, 2, \dots, 2^{|A_j|}\}$ is the set of all subsets of $A_J$ except $\emptyset$, $\boldsymbol{\tau}_k$ is a sub-matrix, which can be obtained by removing, from the $K \times K$ identity matrix, those rows that are not elements of $E_{j_k}$, $\boldsymbol{\Sigma}_k$ is a sub-matrix of $\boldsymbol{\Sigma}$ with rows and columns that are elements of  $E_{j_k}$, $D_K$ is a $K^2 \times(K(K+1)/2)$ duplication matrix.   

Parameter of interest is $\boldsymbol{\eta}=(\lambda_1\mu_1, \dots, \lambda_K\mu_K)$. $\frac{\partial\boldsymbol{\eta}}{\partial\mu_k}=\lambda_k$, $\frac{\partial\boldsymbol{\eta}}{\partial\lambda_k}=\mu_k$.
Thus, $V(\hat{\boldsymbol{\eta}})$ the asymptotic covariance matrix for $\hat{\boldsymbol{\eta}}$, the maximum likelihood estimator of $\boldsymbol{\eta}$ is as follows:
\begin{equation} \label{eq:3.3.5}
    V(\hat{\boldsymbol{\eta}})=CI^{-1}(\boldsymbol{\theta}; P)C^T,
\end{equation}
 where $  	C=\begin{bmatrix}
    C_1  & C_2 & C_3
    \end{bmatrix}$, $C_1=diag(\exp(\mu_1+\Sigma_{11}/2), \dots, \exp(\mu_K+\Sigma_{KK}/2))$, $C_2=diag(\lambda_1 \exp(\mu_1+\Sigma_{11}/2), \dots, \lambda_K \exp(\mu_K+\Sigma_{KK}/2))$, and $C_3$ is a $K \times (K(K+1)/2)$ matrix such that $[C_3]_{ij}=(0.5)\lambda_i \exp(\mu_i+\Sigma_{ii}/2)$ if $j=1+(K-1)i$, and $0$ otherwise. $A(\boldsymbol{\theta}, P)$, the A-optimality criterion for the ZMVLN model is as follows:
\begin{equation} \label{eq:3.3.6}
A(\boldsymbol{\theta}, P)=tr\left(V(\hat{\boldsymbol{\eta}})\right)=tr \left(CI^{-1}(\boldsymbol{\theta}; P)C^T\right).
\end{equation}
    $A(\boldsymbol{\theta}, P)$ is a function of $\boldsymbol{\theta}$. So, similar to the MVN case, we additionally assume that we can use data from a pilot study to obtain a preliminary estimator of $\boldsymbol{\theta}$ and minimize $A(\boldsymbol{\theta}, P)$ after plugging in the estimator. A-OSQD, Bayesian A-OSQD, and mini-max A-OSQD can be defined the same way as in section 3.2.

\subsection{Theoretical Results}
Realistically, some questions are highly correlated. One can assume that there are groups of items such that within-group questions are highly correlated, whereas between-group questions are not highly correlated. Under the assumption, it would be useful to explore the behavior of A-OSQD for different combinations of with-in and between groups correlations.
Suppose we have two groups of questions, where each group has $q$ questions. Suppose that the within-group correlations are equal to $\rho_1$, and between-group correlations are equal to $\rho_2$. Assume that $0<|\rho_2|<|\rho_1|<1$.
Then, $\boldsymbol{\Sigma}$ is as follows:\\
\begin{equation} \label{eq:3.4.1}
    \boldsymbol{\Sigma}=\begin{bmatrix}
    (1-\rho_1)\boldsymbol{I}_q+\rho_1 \boldsymbol{J}_q  & \rho_2 \boldsymbol{J}_q \\
    \rho_2 \boldsymbol{J}_q & (1-\rho_1)\boldsymbol{I}_q+\rho_1 \boldsymbol{J}_q 
    \end{bmatrix},
\end{equation}
where $\boldsymbol{I}_q$ is $q \times q$ identity matrix, $\boldsymbol{J}_q$ is $q \times q$ matrix of one's. Suppose we choose 2 questions for SQD, then
\begin{equation} \label{eq:3.4.2}
  \Sigma_{\{i,j\}}^{-1} = \begin{cases}
\frac{1}{1-\rho_1^2} \begin{bmatrix}
    1  & -\rho_1 \\
    -\rho_1 & 1 
    \end{bmatrix} & \text{  ~~~~~if i, j are in the same group}, \\ \frac{1}{1-\rho_2^2} \begin{bmatrix}
    1  & -\rho_2 \\
    -\rho_2 & 1 
    \end{bmatrix} &\text{  ~~~~~if i, j are in different groups}.
\end{cases}  \\
\end{equation}
By symmetry, we additionally assume that $\pi_{\{i,j\}}=\frac{\pi}{q(q-1)}$ for $i,j$ in the same group, $\frac{1-\pi}{q^2}$ otherwise, where $0 \le \pi \le 1$. We aim to find an optimal $pi$ that minimize the A-optimality criterion. Under the MVN model, using (\ref{eq:4}), (\ref{eq:3.4.1}), and (\ref{eq:3.4.2}), one can derive that
\begin{equation} \label{eq:3.4.3}
    	I(\boldsymbol{\mu}; \pi)=\begin{bmatrix}
    a\boldsymbol{I}_q+b \boldsymbol{J}_q &  c\boldsymbol{J}_q\\
    c\boldsymbol{J}_q & a\boldsymbol{I}_q+b \boldsymbol{J}_q
    \end{bmatrix},
\end{equation}
where 
\begin{align*}
    a &=\frac{\pi}{q(1-\rho_1^2)}+\frac{1-\pi}{q(1-\rho_2^2)}-b,\\
    b &=-\frac{\pi\rho_1}{q(q-1)(1-\rho_1^2)}, \\
    c &=-\frac{(1-\pi)\rho_2}{q^2(1-\rho_2^2)}.
\end{align*}
We define $A(\pi)=tr(I(\mu ; \pi)^{-1})$, the A-optimality criterion. One can derive $A(\pi)$ as follows:
\begin{align*}
    A(\pi)&=\frac{2q-2}{\frac{\pi}{q(1-\rho_1^2)}+\frac{1-\pi}{q(1-\rho_2^2)}+\frac{\pi\rho_1}{q(q-1)(1-\rho_1^2)}}+\frac{2\left(\frac{\pi(1-\rho_1)}{q(1-\rho_1^2)}+\frac{1-\pi}{q(1-\rho_2^2)}\right)}{\left(\frac{\pi(1-\rho_1)}{q(1-\rho_1^2)}+\frac{1-\pi}{q(1-\rho_2^2)}\right)^2-\frac{(1-\pi)^2\rho_2^2}{q^2(1-\rho_2^2)^2}}.
\end{align*}
Thus, the A-OSQD, the minimizer of $A(\pi)$ is as follow:
\begin{equation} \label{eq:3.4.5}
    \pi_{opt}=arg \min_{\pi \in [0,1]} A(\pi)
\end{equation}
 Under the assumptions above, we can prove the following theorem.
\begin{theorem} \label{thm:1}
As $q$ goes to $+\infty$, $\pi_{opt}$ converges to 1. 
\end{theorem}
\begin{proof}
\begin{align*}
    \frac{1}{q^2}A(\pi)&=\frac{2-2/q}{\frac{\pi}{(1-\rho_1^2)}+\frac{1-\pi}{(1-\rho_2^2)}+\frac{\pi\rho_1}{(q-1)(1-\rho_1^2)}}+\frac{2\left(\frac{\pi(1-\rho_1)}{q(1-\rho_1^2)}+\frac{1-\pi}{q(1-\rho_2^2)}\right)}{\left(\frac{\pi(1-\rho_1)}{(1-\rho_1^2)}+\frac{1-\pi}{(1-\rho_2^2)}\right)^2-\frac{(1-\pi)^2\rho_2^2}{(1-\rho_2^2)^2}}.\\
    \lim_{q \rightarrow \infty} \frac{1}{q^2}A(\pi) &=\frac{2}{\pi\left(\frac{1}{(1-\rho_1^2)}-\frac{1}{(1-\rho_2^2)}\right)+\frac{1}{(1-\rho_2^2)}}.
\end{align*}
The limit of $\frac{1}{q^2}A(\pi)$ is a increasing function of $\pi$, when $|\rho_1|>|\rho_2|$.\\
Thus, $\pi_{opt}$ converges to 1, as $q$ goes to $+\infty$.
\end{proof}

 This result implies that as $q$ increases, the sample of questions from the same group obtains more information than sampling questions from different groups. Denote $P_{SRS}$ the design that chooses questions using SRS. Under the above assumptions,
\[
    \pi_{SRS} \equiv \frac{2 {\binom{q}{2}}}{\binom{2q}{2}}=\frac{q-1}{2q-1}.
\]
It would be interesting to explore the relative efficiency of A-OSQD, which can be defined as $A(\pi_{SRS})/A(\pi_{opt})$. Under the above assumptions, we can prove the following theorem.

\begin{theorem} \label{thm:2}
As $q$ goes to $+\infty$, $A(\pi_{SRS})/A(\pi_{opt})$ converges to $\frac{2/(1-\rho_1^2)}{1/(1-\rho_1^2)+1/(1-\rho_2^2)}$. 
\end{theorem}
\begin{proof}
\begin{eqnarray*}
     \frac{1}{q^2}A(\pi_{SRS}) &=&\frac{2-2/q}{\frac{q-1}{(2q-1)(1-\rho_1^2)}+\frac{q}{(2q-1)(1-\rho_2^2)}+\frac{\rho_1}{(2q-1)(1-\rho_1^2)}}     \\
 &    &+\frac{2\left(\frac{(q-1)(1-\rho_1)}{q(2q-1)(1-\rho_1^2)}+\frac{1}{(2q-1)(1-\rho_2^2)}\right)}{\left(\frac{\pi(1-\rho_1)}{(1-\rho_1^2)}+\frac{1-\pi}{(1-\rho_2^2)}\right)^2-\frac{(1-\pi)^2\rho_2^2}{(1-\rho_2^2)^2}}.\\
    \lim_{q \rightarrow \infty} \frac{1}{q^2}A(\pi_{SRS}) &=&\frac{2}{\frac{1}{2(1-\rho_1^2)}+\frac{1}{2(1-\rho_2^2)}}=\frac{4}{\frac{1}{1-\rho_1^2}+\frac{1}{1-\rho_2^2}}.\\
   \lim_{q \rightarrow \infty} \frac{1}{q^2}A(\pi_{opt}) &=& \frac{2}{\frac{1}{1-\rho_1^2}}.
\end{eqnarray*}
Thus, $A(\pi_{SRS})/A(\pi_{opt})$ converges to $\frac{2/(1-\rho_1)^2}{1/(1-\rho_1^2)+1/(1-\rho_2^2)}$, as $q$ goes to $+\infty$.
\end{proof}

These results imply that the relative efficiency of A-OSQD and SRS depends on the number of questions in a group. As the number of questions in a group increases, the relative efficiency converges to its upper bound. Furthermore, the limit point is a function of $|\rho_1|$ and $|\rho_2|$. If the within-group correlation is larger than the between-group correlation, i.e., \ $|\rho_2|>|\rho_1|$,  the limit is greater than 1. For the larger $|\rho_1|$ and the smaller $|\rho_2|$, the limit of the relative efficiency is larger. Thus, we would conclude that A-OSQD is more efficient when the within-group correlation is high and the between-group correlation is low.\\
Since the values of parameters are unknown in practice, we may plug-in its consistent estimators instead. The following theorems are to  justify the idea.
\begin{theorem} \label{thm:3}
Let $\hat{\pi}_{opt}=\arg \min_{\pi} A(\pi, \hat{\rho}_1, \hat{\rho}_2)$, where $\hat{\rho}_1, \hat{\rho}_2$ are $\sqrt{n}$-consistent estimators of $\rho_1$ and $\rho_2$ respectively, where $n$ is the sample size of preliminary survey data. As $n$ and $q$ go to $+\infty$, $\hat{\pi}_{opt}$ converges to 1. 

\end{theorem}
\begin{proof}
Let $G_n : [0,1] \rightarrow {\rm I\!R}
$ be a function such that $G_n(\pi)=A(\pi, {\rho}_1, {\rho}_2)-A(\pi, \hat{\rho}_1, \hat{\rho}_2), \forall \pi \in [0,1]$. Then, since $A(\pi, {\rho}_1, {\rho}_2)$ is a uniform continuous function of $\rho_1$ and $\rho_2$ and $\hat{\rho}_1, \hat{\rho}_2$ are $\sqrt{n}$-consistent estimators, it can be shown that $G_n \rightarrow 0$ in probability and $\lim \sup_n P(\sup_{\pi} \sup_{\pi' \in B(\pi, \delta)} |G_n(\pi)-G_n({\pi'})|>\epsilon)<\epsilon$.
Thus, by the generic uniform convergence theorem \citep{andrews1992generic}, $\sup_{\pi} |G_n(\pi)| \rightarrow 0$ in probability. Thus, $\hat{\pi}_{opt} \rightarrow \pi_{opt}$ as $n \rightarrow \infty$, for each $q$. As $q \rightarrow \infty$, $\hat{\pi}_{opt} \rightarrow 1$.
\end{proof}

\begin{theorem} \label{thm:4}
As $q$ goes to $+\infty$, and $n$ goes to $+\infty$, $A(\pi_{SRS})/A(\hat{\pi}_{opt})$ converges to $\frac{2/(1-\rho_1^2)}{1/(1-\rho_1^2)+1/(1-\rho_2^2)}$. 

\end{theorem}
\begin{proof}
Since $\hat{\pi}_{opt}$ converges to 1, the result is followed by Theorem \ref{thm:2}.\\
\end{proof}
Theorem \ref{thm:3} and \ref{thm:4} state that although the value of the parameter is unknown, A-OSQD can be found using the A-optimality criterion plugged-in by $\sqrt{n}$-consistent estimator, and the performance of the design will not be significantly damaged.
\section{Simulation Results}

\subsection{Simulation 1: MVN model}
To test the theoretical results, we conduct several simulation studies. We generate $\boldsymbol{Y_j}\sim MVN(\boldsymbol{\mu}, \boldsymbol{\Sigma})$, for $\boldsymbol{j}=1, \dots, 100,000$, where 
	$\boldsymbol{\mu} =(\boldsymbol{1}_q^T, 2\boldsymbol{1}_q^T, \dots, g\boldsymbol{1}_q^T)^T$, \[\boldsymbol{\Sigma}= \begin{bmatrix}
    (1-\rho_1)I_q+\rho_1 \boldsymbol{J}_q  & \rho_2 \boldsymbol{J}_q & \dots & \rho_2 \boldsymbol{J}_q \\
    \rho_2 \boldsymbol{J}_q & (1-\rho_1)I_q+\rho_1 \boldsymbol{J}_q  & \dots & \rho_2 \boldsymbol{J}_q \\
    \vdots & \vdots & \ddots&  \vdots & \\
    \rho_2 \boldsymbol{J}_q & \rho_2 \boldsymbol{J}_q & \dots & (1-\rho_1)I_q+\rho_1 \boldsymbol{J}_q 
    \end{bmatrix},\]

     and $\boldsymbol{1}_q$ is the $q \times 1$ vector of ones. $g$ is the number of groups, and $q$ is the number of questions in a group. The data is regarded as population. 1,000 Monte Carlo (MC) samples with size $n=1,000$ are sampled from the population using SRS. For each MC sample, we apply two SQDs: 
   \begin{enumerate}
   \item SRS.
   \item A-OSQD based on the true covariance structure.
   \end{enumerate}
   For each respondent, 2 question items are chosen out of $gq$ question items using one of the SQDs. The responses are observed for the items chosen by each design, missing otherwise. Therefore, we can obtain two different observed data set for other designs. For each observed data, we use the EM-algorithm to compute the MLE of $\boldsymbol{\mu}$. The “norm” package of R Core Team \citep{novo2003norm, cranr} is used to implement the EM-algorithm. The two designs are compared using the sum of MSEs and the A-optimality criterion. The A-optimality criterion is computed based on (\ref{eq:6}). The sum of MSEs is computed as follows:
   \begin{equation} \label{eq:4.1.1}
       MSE=\sum_{i=1}^{1000}\sum_{j=1}^{qg} \left( \hat{\mu}_j^{i}-\mu_j\right)^2/(1000qg),
   \end{equation}
   where $\hat{\mu}_j^{i}$ is the MLE of $j$th population mean based on the $i$th MC sample.
   \begin{figure}[h]  
   \includegraphics[scale=0.4]{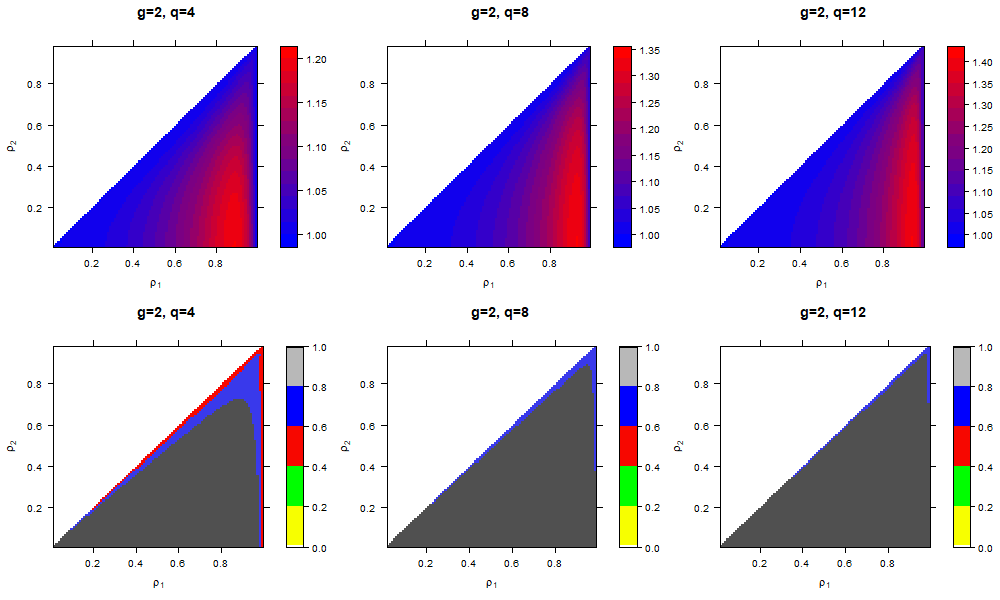}
   \caption{Plots of designs and their relative efficiencies. The first row of plots presents relative efficiencies of SRS design and A-OSQD for all $(\rho_1, \rho_2)$. The second row of plots are $\pi_{\text{A-opt}}(\rho_1, \rho_2)$. For all plots, the number of groups, $g=2$.}
   \label{fig:4.1}
   \end{figure}
    \begin{figure}[h]
   \includegraphics[scale=0.4]{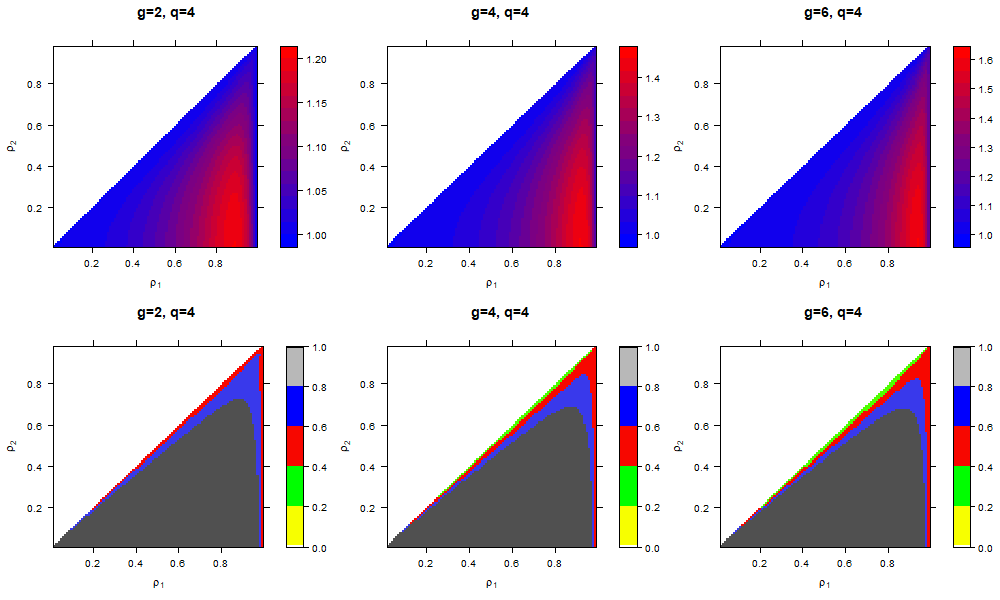}
   \caption{Plots of designs and their relative efficiencies. The first row of plots presents relative efficiencies of SRS design and A-OSQD for all $(\rho_1, \rho_2)$. The second row of plots are $\pi_{\text{A-opt}}(\rho_1, \rho_2)$. For all plots, the number of question items in a group, $q=4$.}
    \label{fig:4.2}
   \end{figure}
   Figure \ref{fig:4.1} and \ref{fig:4.2} present plots of designs and their relative efficiencies. The first row of plots presents relative efficiencies of SRS design and A-OSQD for all $(\rho_1, \rho_2)$. The second row of plots presents $\pi_{\text{A-opt}}(\rho_1, \rho_2)$, which is the sum of A-OSQD probability of choosing two questions from the same group, defined in (\ref{eq:3.4.5}). For all plots in Figure \ref{fig:4.1}, $g$, the number of groups is two, and $q$, the number of question items in a group is increased. For all plots in Figure \ref{fig:4.2}, $q$, the number of question items in a group is two, and $g$, the number of groups is increased. When the number of question items increases, the maximum relative efficiency increases. In Figure \ref{fig:4.1}, the second row of plots shows that the pattern of A-OSQD changes as the number of questions in a group increases. But, the pattern of A-OSQD remains the same in Figure \ref{fig:4.2}.
   Table \ref{table:3} presents the sum of MSEs and the A-optimality criterion of SRS and A-OSQD when $\rho_1=0.8$ and $\rho_2=0.4$. $RE_{MSE}$ is the relative efficiency of A-OSQD and the SRS design with respect to the sum of MSE, and $RE_{A}$ is the relative efficiency of two designs with respect to the A-optimality criterion. $RE_{MSE}$ and $RE_{A}$ have similar values for all rows indicating that the sample size of 1,000 is large enough for MLEs to achieve its asymptotic normality. It shows that the relative efficiencies are greater than 1, indicating that the A-OSQD leads to more efficient estimators than the SRS design. It can also be seen that the relative efficiency increases as either the number of questions in a group or the number of groups increases. We have tried several combinations of $(\rho_1, \rho_2)$, and obtained similar results.

\begin{table}[ht]
\caption{The sums of MSEs and the A-optimality criterion of two designs from simulation 1 when $\rho_1=0.8$ and $\rho_2=0.4$}

\centering
\begin{tabular}{rrrrrrrrrr}
  \hline
$g$ & $q$  & $MSE_{SRS}$ & $MSE_{opt}$ & $RE_{MSE}$ & $A_{SRS}$ & $A_{opt}$ & $RE_{A}$ \\ 
  \hline
2& 4 & 0.0026 & 0.0025 & 1.0454 & 20.5173 & 19.6397 & 1.0447 \\ 
2&   8 & 0.0046 & 0.0041 & 1.1390 & 73.3262 & 64.5480 & 1.1360 \\ 
 2&  12 & 0.0048 & 0.0039 & 1.2276 & 111.5742 & 91.8240 & 1.2151 \\ 
   \hline
   2 & 4 &  0.0026 & 0.0025 & 1.0454 & 20.5173 & 19.6397 & 1.0447 \\ 
 4 & 4 & 0.0057 & 0.0049 & 1.1645 & 89.8490 & 77.3975 & 1.1609 \\ 
 6 & 4 & 0.0077 & 0.0059 & 1.3090 & 180.0130 & 139.0093 & 1.2950 \\ 
    \hline
\end{tabular}
\caption*{Notes: $MSE_{SRS}$, the sum of MSEs of SRS; $MSE_{opt}$, the sum of MSEs of A-OSQD; $RE_{MSE}=MSE_{SRS}/MSE_{opt}$, the relative efficiency of two designs with respect to MSE; $A_{SRS}$, A-optimality criterion of SRS; $A_{opt}$, A-optimality criterion of A-OSQD; $RE_{A}=A_{SRS}/A_{opt}$, the relative efficiency of two designs with respect to A-optimality criterion.}
\label{table:3}
\end{table}


\subsection{Simulation 2: ZMVLN model}
For the ZMVLN model, we also conduct several simulation studies. In scenario 3, we increased the number of questions in a group when the number of groups is fixed. We generat $\boldsymbol{Y_j}\sim ZMVLN(\boldsymbol{\mu}, \boldsymbol{\Sigma}, \boldsymbol{\lambda})$, for $\boldsymbol{j}=1, \dots, 100,000$, where 
	$\boldsymbol{\mu} =(\boldsymbol{1}_q^T, 2\boldsymbol{1}_q^T)^T$,
	\[
	\boldsymbol{\Sigma}= \begin{bmatrix}
    (1-\rho_1)\boldsymbol{I}_q+\rho_1 \boldsymbol{J}_q  & \rho_2 \boldsymbol{J}_q \\
    \rho_2 \boldsymbol{J}_q & (1-\rho_1)\boldsymbol{I}_q+\rho_1 \boldsymbol{J}_q
    \end{bmatrix},
    \] and $\boldsymbol{\lambda}=(p_1 \boldsymbol{1}_{q}^T, p_2 \boldsymbol{1}_{q}^T)^T$.
    This simulation scenario covers two cases related to the probability of non-zero responses: the same probability case and the different probability case. In the same probability case, we generated data from ZMVLN with the same probability of non-zero responses, i.e.\, $p_1=p_2=0.8$.
In the different probability case, probabilities of non-zero responses are different for questions, i.e.\, $p_1=0.8$ and $p_2=0.6$. We regarded the data as population. 1,000 Monte Carlo (MC) samples with size $n$ are sampled from the population using SRS. For each MC sample, we apply two SQDs: SRS and A-OSQD. Two questions are chosen out of $2m$ questions for each respondent using one of the two designs. For each SQD, we obtain an observed data set. For each observed data, we use the EM-algorithm to compute the MLE of $\boldsymbol{\mu}$ and $\boldsymbol{\Sigma}$. For observed non-zero responses, we apply the exponential transform. Regard the data with the transformed responses only as the observed data set of MVN distribution. Then, we implement the EM-algorithm using The “norm” package of R Core Team \cite{norm, cranr}. $\hat{\lambda}_k$, the estimator of $\lambda_k$ for $k=1, 2, \dots, K$ is computed as follows:
\begin{equation} \label{eq:4.2.1}
    \hat{\lambda}_k=\frac{\sum_{i=1}^n I(\text{$Y_{ik}$ is observed and non-zero})}{\sum_{j=1}^n I(\text{$Y_{jk}$ is observed})}
\end{equation}
$\hat{\eta}_k$, the estimator of $k$th population mean for $k=1, 2, \dots, K$ is computed as follows:
\begin{equation} \label{eq:4.2.2}
    \hat{\eta}_k=\hat{\lambda}_k \hat{\mu}_k.
\end{equation}
    The two designs are compared using the sum of MSEs and the A-optimality criterion. The A-optimality criterion is computed based on (\ref{eq:3.3.6}). The sum of MSEs is computed as follows:
   \begin{equation} \label{eq:4.2.3}
       MSE=\sum_{i=1}^{1000}\sum_{j=1}^{qg} \left( \hat{\eta}_j^{i}-\eta_j\right)^2/(1000qg)
   \end{equation}

We also calculate the relative efficiency of the two designs and jackknife standard errors (SE) \citep{efron1994introduction} of the relative efficiency to evaluate its variability.
Table \ref{table:4} presents the sum of MSEs of SRS and A-OSQD, the relative efficiency of the two SQDs based on the A-optimality criterion, and the sum of MSEs, and the jackknife SE of the relative efficiency when $\rho_1=0.8$ and $\rho_2=0.2$. $RE_A$, the asymptotic relative efficiency, and $RE_{MSE}$, the relative efficiency based on simulation MSE have some differences, but the differences decrease as the sample sizes increase. When $gq$ the total number of question items is small, $RE_A$ and $RE_{MSE}$ have similar values when $n=3000$. But, as $gq$ increases, the difference between those two increases. This result indicates that as $gq$ increases, a larger sample size is required for the asymptotic normality of the MLE. Similar to the results of the MVN model, both $RE_A$ and $RE_{MSE}$ increase as either the number of questions in a group or the number of groups increases. Also, the relative efficiencies are greater than 1, indicating the A-OSQD leads to more efficient estimators than the SRS design.

\begin{table}[H]

\centering
\caption{The sums of MSEs of two designs, the relative efficiencies, and the jackknife SE of the relative efficiency from the simulation 2 when $\rho_1=0.8$ and $\rho_2=0.2$}
\begin{tabular}{rrrrrrrrrr}
  \hline
$p_1$ & $p_2$ & g & q & n & $RE_{A}$ & $MSE_{SRS}$ & $MSE_{opt}$ & $RE_{MSE}$ & $SE_{RE}$  \\ 
  \hline
0.8 & 0.8 &  2  & 4 & 2000 & 1.0981 & 0.1765 & 0.1771 & 0.9969 & 0.0284 \\ 
& & 2  & 4 & 3000 & 1.0981 & 0.1201 & 0.1100 & 1.0913 & 0.0295 \\ 
& & 2  &8 & 2000 & 1.1428 & 0.3581 & 0.2954 & 1.2121 & 0.0250 \\ 
& & 2  & 8 & 3000 & 1.1428 & 0.2307 & 0.1977 & 1.1669 & 0.0235 \\ 
& & 2  &12 & 2000 & 1.1574 & 0.5128 & 0.4315 & 1.1884 & 0.0207 \\ 
& & 2  & 12 & 3000 & 1.1574 & 0.3417 & 0.2880 & 1.1865 & 0.0210 \\ 
   \hline
0.8 & 0.6 & 2  &4 & 2000 & 1.0730 & 0.1603 & 0.1543 & 1.0387 & 0.0293 \\ 
& & 2  &  4 & 3000 & 1.0730 & 0.1091 & 0.1018 & 1.0710 & 0.0305 \\ 
& & 2  &8 & 2000 & 1.1009 & 0.3272 & 0.2927 & 1.1178 & 0.0224 \\ 
& &  2  & 8 & 3000 & 1.1009 & 0.2102 & 0.1847 & 1.1383 & 0.0215 \\ 
& & 2  &12 & 2000 & 1.1095 & 0.4779 & 0.4082 & 1.1707 & 0.0193 \\ 
& &  2  & 12 & 3000 & 1.1095 & 0.3164 & 0.2768 & 1.1429 & 0.0191 \\ 
   \hline
0.8 & 0.8 & 2 &4 & 2000 & 1.0981 & 0.1765 & 0.1771 & 0.9969 & 0.0284 \\ 
& &   2 &4 & 3000 & 1.0981 & 0.1201 & 0.1100 & 1.0913 & 0.0295 \\ 
   
& & 4 &4 & 2000 & 1.1854 & 0.4065 & 0.3434 & 1.1836 & 0.0255 \\ 
 & &  4 &4 & 3000 & 1.1854 & 0.2691 & 0.2257 & 1.1924 & 0.0255 \\ 
& & 6 &4 & 2000 & 1.2210 & 0.6439 & 0.5031 & 1.2799 & 0.0226 \\ 
& &   6 &4 & 3000 & 1.2210 & 0.4156 & 0.3380 & 1.2296 & 0.0226 \\
   \hline
0.8 & 0.6 & 2 &4 & 2000 & 1.0730 & 0.1603 & 0.1543 & 1.0387 & 0.0293 \\ 
 & &   2 &4 & 3000 & 1.0730 & 0.1091 & 0.1018 & 1.0710 & 0.0305 \\ 
 & &  4 &4 & 2000 & 1.1403 & 0.3709 & 0.3070 & 1.2080 & 0.0251 \\ 
 & &   4 &4 & 3000 & 1.1403 & 0.2360 & 0.2034 & 1.1603 & 0.0236 \\ 
 & & 6 &4 & 2000 & 1.1666 & 0.5567 & 0.4572 & 1.2177 & 0.0208 \\ 
 & &   6 &4 & 3000 & 1.1666 & 0.3671 & 0.3031 & 1.2112 & 0.0209 \\  
   \hline
\end{tabular}
\caption*{Notes: $MSE_{SRS}$, the sum of MSEs of SRS; $MSE_{opt}$, the sum of MSEs of A-OSQD; $RE_{MSE}=MSE_{SRS}/MSE_{opt}$, the relative efficiency of two designs with respect to MSE; $RE_{A}=A_{SRS}/A_{opt}$, the relative efficiency of two designs with respect to A-optimality criterion where $A_{SRS}$ is the A-optimality criterion of SRS and $A_{opt}$ is the A-optimality criterion of A-OSQD; $SE_{RE}$, the jackknife SE of the $RE_{MSE}$.}
\label{table:4}
\end{table}

\subsection{Simulation 3: Comparison with the deterministic question order}

\cite{zhang2019active} proposed the active question selection method that obtained a deterministic question order for all respondents for the Gaussian response model. To compare the A-OSQD with the active question selection method in \cite{zhang2019active}, we conduct simulation studies. As in simulation 1, we generate MVN responses of size 100,000 and regard it as population. We repeat the following procedure 1000 times using the simulated population, so we have 1,000 Monte Carlo (MC) samples. We construct two simulation setups. In setup 1, we apply five designs: 
   \begin{enumerate}
   \item SRS: Draw a sample of size $n=100*K/2$, where $K=gq$ is the total number of questions. Choose two questions using SRS for each respondent.
   \item A-OSQD based on the true covariance structure: Draw a sample of size $n= (100K)/2$. Choose two questions using the A-OSQD for each respondent.
   \item Deterministic question order 1 : Draw a sample of size $n= 50+(50K)/2$. For 50 respondents, observe all variables. For $(50K)/2$ respondents, observe the first variable in the first group and the last variable in the last group.
    \item Deterministic question order 2 : Draw a sample of size $n= 50+(50K)/2$. For 50 respondents, observe all variables. For $(50K)/2$ respondents, observe the first two variables in the first group.
    \item Full : Draw a sample of size $n=100$. Observe all variables.
   \end{enumerate}
   We use SRS to draw a sample from the simulated population for all designs. Also, we can note that the numbers of observed responses are equal to $100K$ for all designs. 
  For SRS, A-OSQD and Full, we use the EM-algorithm to compute the MLE of $\boldsymbol{\mu}$. For the two deterministic question ordering designs, we use the multiple imputation to estimate $\boldsymbol{\mu}$, because the method of \cite{zhang2019active} is based on multiple imputation. We generate the imputed values using the conditional distribution of missing variables given observed variables where the parameters are plugged-in by the estimators based on the observed data set. Then, the column means of imputed data set are the estimators of the population means. The five designs are compared using the sum of MSEs.
   Table \ref{table:5.1} presents the sum of MSEs of the five designs from setup 1, when $\rho_1=0.8$ and $\rho_2=0.2$. Even though the five designs have the same number of observed responses, the A-OSQD has the smallest sum of MSEs. Therefore, the A-OSQD is more efficient than the others in aggregate MSE. We can also note that the two deterministic question order designs have larger aggregate MSEs than that of the SRS design.

\begin{table}[H]
\caption{The sums of MSEs of five designs from the setup 1 of simulation 3 when $\rho_1=0.8$ and $\rho_2=0.2$}

\centering
\begin{tabular}{rrrrrrr}
  \hline
$g$ & $q$ & SRS & A-OSQD & Det1 & Det2 & full \\ 
  \hline
2 & 4 & 7.15 & 6.51 & 8.37 & 13.81 & 9.91 \\ 
2 & 8  & 6.68 & 5.56 & 8.31 & 13.55 & 10.11 \\ 
2 & 12  & 6.43 & 4.91 & 7.95 & 13.30 & 10.10 \\ 
   \hline
2 & 4 & 7.15 & 6.51 & 8.37 & 13.81 & 9.91 \\ 
  4 & 4 & 8.24 & 6.54 & 13.70 & 16.55 & 10.22 \\ 
  6 & 4 & 8.99 & 6.76 & 15.66 & 17.61 & 9.87 \\ 
   \hline

\end{tabular}
\caption*{Notes: The last five columns of the table presents the sum of MSEs of each design. Det1 is the deterministic question order 1, and Det2 is the deterministic question order 2}
\label{table:5.1}
\end{table}

In setup 2, we draw two samples from the simulated population. The first sample of size 100 is regarded as the previous survey. We observe all variables for the first sample. The first half of variables in the first sample are shifted by -0.2, while the other half of variables in the sample are shifted by 0.2. Thus, the first data and second data have a different population mean. Then, the second sample of size $n$ is regarded as the current survey. We apply the following five designs to the second sample: 
   \begin{enumerate}
   \item SRS: The size of the second sample is  $n=100*K/2$, where $K=gq$ is the total number of questions. Choose two questions using SRS for each respondent.
   \item A-OSQD based on the true covariance structure: The size of the second sample is  $n=100*K/2$. Choose two questions using the A-OSQD for each respondent.
   \item Deterministic question order 1: The size of the second sample is  $n=100*K/2$. Observe the first variable in the first group and the last variable in the last group.
    \item Deterministic question order 2: The size of the second sample is  $n=100*K/2$. Observe the first two variables in the first group.
    \item Full : Draw a sample of size $n=100$. Observe all variables.
   \end{enumerate}

Using the first sample data, we find the A-OSQD. Then, we compute the MLE of $\boldsymbol{\mu}$ for SRS, A-OSQD, and Full design. For the two deterministic question order designs, we use the first sample data to compute the imputed values. The imputed values are generated from the conditional distribution of missing variables given observed variables after the parameters are plugged-in by the estimators based on the first data. Then, the column means of the imputed data set are the estimators of the population means. The five designs are compared using the sum of MSEs. Table \ref{table:5.1} presents the sum of MSEs of the five designs from setup 2, when $\rho_1=0.8$ and $\rho_2=0.2$. Because the correlation structures of the first and second data are the same, the A-OSQD performs better than the others in aggregate MSE. Since the first and second data have a different population mean, the imputation procedure can be ruined. For this reason, the two deterministic question order designs perform worse than the A-OSQD in terms of the aggregate MSE. Thus, when the population means are different for the previous survey and the current survey, but the correlation structures are the same, the A-OSQD is a more efficient design than the other four designs.
\begin{table}[H]
\caption{The sums of MSEs of five designs from the setup 2 of simulation 3 when $\rho_1=0.8$ and $\rho_2=0.2$}

\centering
\begin{tabular}{rrrrrrr}
  \hline
$g$ & $q$ & SRS & A-OSQD & Det1 & Det2 & full \\ 
  \hline
2 & 4 & 7.30 & 6.53 & 6.62 & 36.93 & 10.20 \\ 
2 & 8 & 6.61 & 5.32 & 6.46 & 36.73 & 9.61 \\ 
2 & 12 & 6.27 & 4.92 & 6.81 & 37.35 & 9.73 \\ 
   \hline
   2 & 4 & 7.30 & 6.53 & 6.62 & 36.93 & 10.20 \\ 
  4 & 4 & 8.31 & 6.52 & 28.01 & 43.73 & 10.25 \\ 
  6 & 4 & 9.07 & 6.62 & 36.00 & 47.09 & 10.13 \\ 
   \hline

\end{tabular}
\caption*{Notes: The last five columns of the table presents the sum of MSEs of each design. Det1 is the deterministic question order 1, and Det2 is the deterministic question order 2}
\label{table:5.2}
\end{table}

\subsection{Simulation 4: Comparison of locally optimal design and global optimal designs}
   \begin{figure}[h]  
   \includegraphics[scale=0.5]{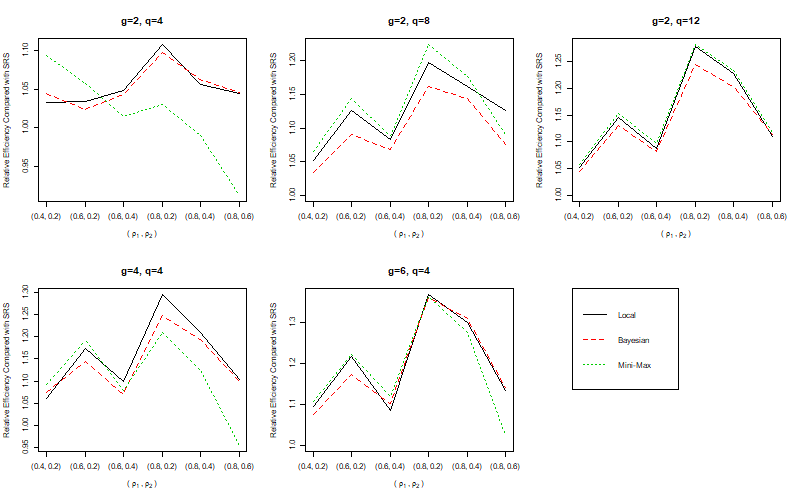}
   \caption{Plots of relative efficiencies of designs compared with SRS design. The first row of plots presents relative efficiencies of SRS design and A-OSQD for all $(\rho_1, \rho_2)$. The second row of plots are $\pi_{\text{A-opt}}(\rho_1, \rho_2)$. For all plots, the number of groups, $g=2$.}
   \label{fig:4.3}
   \end{figure}
This simulation study compares the locally optimal design and global optimal designs. Data generation is similar to simulation 1: MVN model. For six combinations of true $(\rho_1, \rho_2)$: {(0.4, 0.2), (0.6, 0.2), (0.8, 0.2), (0.8, 0.4), (0.8, 0.6)}, we generate population of size 1,000,000 from the MVN model. Then, we draw a simple random sample of size 1,000 from the population. The size of the MC sample is 1,000. For each MC sample, we apply the following four designs:
   \begin{enumerate}
   \item SRS.
   \item A-OSQD based on the true $(\rho_1, \rho_2)$ (Local A-OSQD).
   \item Bayesian A-OSQD where the prior is uniform on the space $0<\rho_2<\rho_1<1$.
   \item Mini-max A-OSQD.
   \end{enumerate}
    A-OSQD, Bayesian A-OSQD, and mini-max A-OSQD are compared in terms of the relative efficiency compared with SRS, where the relative efficiency is the ratio of the sum of simulation MSEs. Figure \ref{fig:4.3} presents the plots of the relative efficiencies for five cases. In most cases, the relative efficiencies of local A-OSQD and Bayesian A-OSQD are similar. However, in some cases, mini-max A-OSQD has substantially different values of relative efficiencies compared with those of local A-OSQD. Thus, when the true values of parameters are unavailable, Bayesian A-OSQD is preferred to mini-max design since it is more stable.


\section{Application to The 2016 PDS Data}
PDS is a nationwide survey for the AVMA. This survey gathers information about the pets of households. The questionnaire consists of questions about the types of pets a household owned (e.g., dogs, cats, and birds), the counts of those pets a household owned, and the amount of money a household paid for a particular type of care (e.g., veterinary clinic, and hospital). This study extracts data from 8,147 respondents from the survey who owned at least one dog or one cat during  2016. We also choose 12 continuous variables about dogs and cats for our study. Table \ref{table:6} presents the information about the 12 variables.

\begin{table}[H]
\caption{} \label{table:6}
\resizebox{\columnwidth}{!}{%
\begin{tabular}{|l|l|}
\hline
Variable & Label                                                                                                                                                                            \\ \hline
qd18     & How much money did you spend at the veterinarian on (all) your dog(s) in total last year?                                                                                        \\ \hline
qd21     & \begin{tabular}[c]{@{}l@{}}How much did you pay in total last year for all the routine check-ups/preventive care for\\  all your dog(s) to the following providers?\end{tabular} \\ \hline
qd25     & How much did you pay in 2016 to have your dog(s) spayed?                                                                                                                         \\ \hline
qd26     & How much did you pay in 2016 to have your dog(s) neutered?                                                                                                                       \\ \hline
qd29     & How much did you pay per day to board one dog?                                                                                                                                   \\ \hline
qd32     & How much did you pay per visit to groom, clip, or clean one dog?                                                                                                                  \\ \hline
qc18     & How much money did you spend at the veterinarian on (all) your cat(s) in total last year?                                                                                        \\ \hline
qc21     & \begin{tabular}[c]{@{}l@{}}How much did you pay in total last year for all the routine check-ups/preventive care for \\ all your cat(s) to the following providers?\end{tabular} \\ \hline
qc25     & How much did you pay in 2016 to have your cat(s) spayed?                                                                                                                         \\ \hline
qc26     & How much did you pay in 2016 to have your cat(s) neutered?                                                                                                                       \\ \hline
qc29     & How much did you pay per day to board one cat?                                                                                                                                   \\ \hline
qc32     & How much did you pay per visit to groom, clip, or clean one cat?                                                                                                                  \\ \hline
\end{tabular}
}
\end{table}

We use this data to examine the performances of local A-OSQD, Bayesian A-OSQD, and mini-max A-OSQD. 
We consider the ZMVLN model because the data variables are positive and non-symmetric with many zeros. First, we split the original data set into two parts: pilot study data of size 4000 and current study data of size 4147. We use the pilot study data to compute  $\hat{\boldsymbol{\theta}}_p=(\hat{\boldsymbol{\mu}}_p,  \hat{\boldsymbol{\Sigma}}_p)$, the preliminary estimates of $\boldsymbol{\theta}=(\boldsymbol{\mu}, \boldsymbol{\Sigma})$ and use them to find three versions of A-OSQD. The four designs that choose two questions from the complete questionnaire in this example are as follows:
\begin{enumerate}
    \item SRS
    \item A-OSQD
    \item Bayes A-OSQD: $P_{B}=\arg\min_P \int A(\pi, \boldsymbol{\theta})f(\boldsymbol{\theta})d\boldsymbol{\theta}$, where $f(\boldsymbol{\theta})$ is a density function of Inverse-Wishart Distribution of parameter $(\hat{\boldsymbol{\mu}}_p, 1, (4,000)\hat{\boldsymbol{\Sigma}}_{p}, 4000)$.
    \item Mini-Max A-OSQD: $P_{MM}=\arg\min_P \max_{\Sigma} A(\pi, \Sigma)$.
\end{enumerate}
We apply the four designs to the current study data and estimate population means based on the incomplete data induced by SQDs. We compare the four designs in terms of the sum of the squared distance to the full data MLE.
We randomly split the data 10 times. We repeat the procedure 1000 times for each split, i.e.\, we have 10,000 MC samples. The estimator from the full data ($n=8147$) is regarded as the true parameter. Then, the sum of MSEs is computed based on four different designs: SRS and A-OSQD. Table \ref{table:7} presents the sum of MSEs of four different designs. The (local) A-OSQD has the smallest sum of MSEs, which coincides with the simulation results. Also, the two global optimal designs: the Bayes design and the mini-max design, have a smaller sum of MSEs than the SRS design. It implies that  global optimal designs can be useful alternatives to the local optimal design in the absence of information on true parameters.

\begin{table}[ht]
\caption{The sums of MSEs of four designs from the real data application} \label{table:7}
\centering
\begin{tabular}{rrrr}
  \hline
$MSE_{SRS}$ & $MSE_{opt}$ & $MSE_{B}$ &  $MSE_{MM}$\\ 
  \hline
363.10 & 323.59 & 344.85 & 329.78 \\
   \hline

\end{tabular}
\caption*{Notes : $MSE_{SRS}$, the sum of MSE of SRS; $MSE_{opt}$, the sum of MSE of (local)  A-OSQD; $RE=  MSE_{SRS}/MSE_{opt}$, the relative efficiency of two designs with respect to MSE}

\end{table}

\section{Conclusions and Discussions}
In the present research, SQDs are approached as probability sampling designs, and the theory of optimal design is used to define the A-OSQD. The A-OSQD for surveys with multiple continuous variables is defined and examined. The feasibility and benefits of the A-OSQD are demonstrated in the simulation studies and the application to the 2016 PDS data. Compared to the SRS design, the A-OSQD performs better for both the MVN model and the ZMVLN model so that it can be useful for surveys with not only symmetric variables but also asymmetric variables or many zero cases. In addition, when the previous and current surveys have a difference in population mean, while the correlation structures are homogeneous, the A-OSQD is more efficient than the other four baselines. In the simulation studies and the example of 2016 PDS data, two global optimal designs: Bayesian design and mini-max design, perform similar to the local A-OSQD, which justify the use of global optimal designs when the values of true parameters are unknown. Moreover, we identify the critical factors for the gain of implementing the A-OSQD: within-group correlations, between-group correlations, and the number of questions. 
Thus, we expect that the proposed SQD methodology can be used for massive surveys as it reduces the loss of information due to using an SQD when the model assumptions are suitable, and the sample size is large enough.

Our approach can also be applied to other optimality criteria. Depending on the purpose of a survey and the types of variables, a suitable optimality criterion can be chosen or developed. One limitation of the A-optimality criterion used in this study is that it depends on the scales of variables. Thus, scale-invariant optimality criteria can be considered alternatives, such as the D-optimality criterion. The following two assumptions are imposed in this study: the survey data has continuous variables only and the sampling method for units (or respondents) is SRS. However, our study can be extended by considering other sampling methods for units and other types of variables, such as ordinal or nominal variables. Such extensions can be interesting topics for future studies.

\bigskip
\begin{center}
{\large\bf SUPPLEMENTARY MATERIAL}
\end{center}

\begin{description}

\item[Title:] Brief description. (file type)

\item[R-package for  MYNEW routine:] R-package ÒMYNEWÓ containing code to perform the diagnostic methods described in the article. The package also contains all datasets used as examples in the article. (GNU zipped tar file)

\item[HIV data set:] Data set used in the illustration of MYNEW method in Section~ 3.2. (.txt file)

\end{description}

\section{BibTeX}

We hope you've chosen to use BibTeX!\ If you have, please feel free to use the package natbib with any bibliography style you're comfortable with. The .bst file agsm has been included here for your convenience. 

\bibliographystyle{agsm}
\bibliography{References}
\end{document}